\documentclass[fleqn,usenatbib]{mnras}

\usepackage{newtxtext,newtxmath}

\usepackage[T1]{fontenc}
\usepackage{ae,aecompl}

\usepackage{graphicx}	
\usepackage{amsmath}	
\usepackage{amssymb}	

\usepackage{times}

\newcommand{\magtwo}{Mg~{\small{II}}}
\newcommand{\ctwo}{C~{\small{II}}}
\newcommand{\cthree}{C~{\small{III}}}
\newcommand{\comm}[1]{\textcolor{black}{#1}}

\title[Fluorescent C II* 1335\AA]{Fluorescent C II* 1335\AA \ emission spectroscopically resolved in a galaxy at $z=5.754$}

\author[S. E. I. Bosman et al.]{
Sarah E. I. Bosman,$^{1}$\thanks{E-mail: ucapeib@ucl.ac.uk}
Nicolas Laporte,$^{1}$
Richard S. Ellis,$^{1}$
Masami Ouchi,$^{2,3}$ \newauthor
Yuichi Harikane$^{2,4}$\\
$^{1}$Department of Physics and Astronomy, University College London, Gower Street, London WC1E 6BT, UK\\
$^{2}$Institute for Cosmic Ray Research, The University of Tokyo, 5-1-5 Kashiwanoha, Kashiwa, Chiba 277-8582, Japan\\
$^{3}$Kavli Institute for the Physics and Mathematics of the Universe (WPI), University of Tokyo, Kashiwa 277-8583, Japan\\
$^{4}$Department of Physics, Graduate School of Science, University of Tokyo, 7-3-1 Hongo, Bunkyo, Tokyo, 113-0033, Japan
}

\date{Accepted 2019 May 24. Received 2019 April 29; in original form 2019 February 11}

\pubyear{2019}

\begin{document}
\label{firstpage}
\maketitle

\begin{abstract}
We report the discovery of the first spectroscopically resolved \ctwo/\ctwo* 1334, 1335 doublet in the Lyman-break galaxy J0215--0555 at $z_{\text{Ly}\alpha}=5.754$. The separation of the resonant and fluorescent emission channels was possible thanks to the large redshift of the source and long integration time, as well as the small velocity width of the feature, $0.6\pm0.2$\AA. We model this emission and find that at least two components are required to reproduce the combination of morphologies of \ctwo* emission, \ctwo \  absorption and emission, and Lyman-$\alpha$ emission from the object. We suggest that the close alignment between the fluorescence and Lyman-$\alpha$ emission could indicate an ionisation escape channel within the object. While the faintness of such a \ctwo/\ctwo* doublet makes it prohibitively difficult to pursue for similar systems with current facilities, we suggest it can become a valuable porosity diagnostic in the era of JWST and the upcoming generations of ELTs.  

\end{abstract}




\section{Introduction}

An enduring puzzle in the study of galaxies at $z>5$ concerns the contribution of these early galaxies to hydrogen reionisation. 
While the commonly held view is that faint early galaxies provide the vast majority of the UV radiation required for this transition, this appears to require the existence of fainter galaxies than the currently accessible limits of $M_{\text{UV}} = -15$, and with either ionisation potentials $\xi_\text{ion}$ or escape fractions $f_\text{esc}$ higher than those of galaxies at later times (see e.g. \citealt{Fan06, Robertson15, Stark16}). 
Some models have indeed predicted enhancements of these parameters at high redshifts caused by the higher star formation rates (SFRs), different stellar populations and dust content, and/or low metallicities of galaxies at $z>5$ (e.g. \citealt{Yajima11, Eldridge12, Fontanot14, Topping15, Stanway16}; but see \citealt{Gnedin08, Ma15}). 
Locally, detailed studies of `analogue' galaxies have linked highly ionising radiation output with compactness and galaxy-scale outflows \citep{Stanway14, Nakajima18, Fletcher18}. Information about the detailed properties of high-$z$ galaxies is therefore now needed, both to constrain models and to determine observational trends. Uncovering the properties of $z>5$ galaxies is a key driver of next-generation observational facilities, such as the \textit{James Webb Space Telescope} \citep{JWST} and the Extremely Large Telescope \citep{ELT}.

\begin{figure*}
	\includegraphics[width=0.9\textwidth]{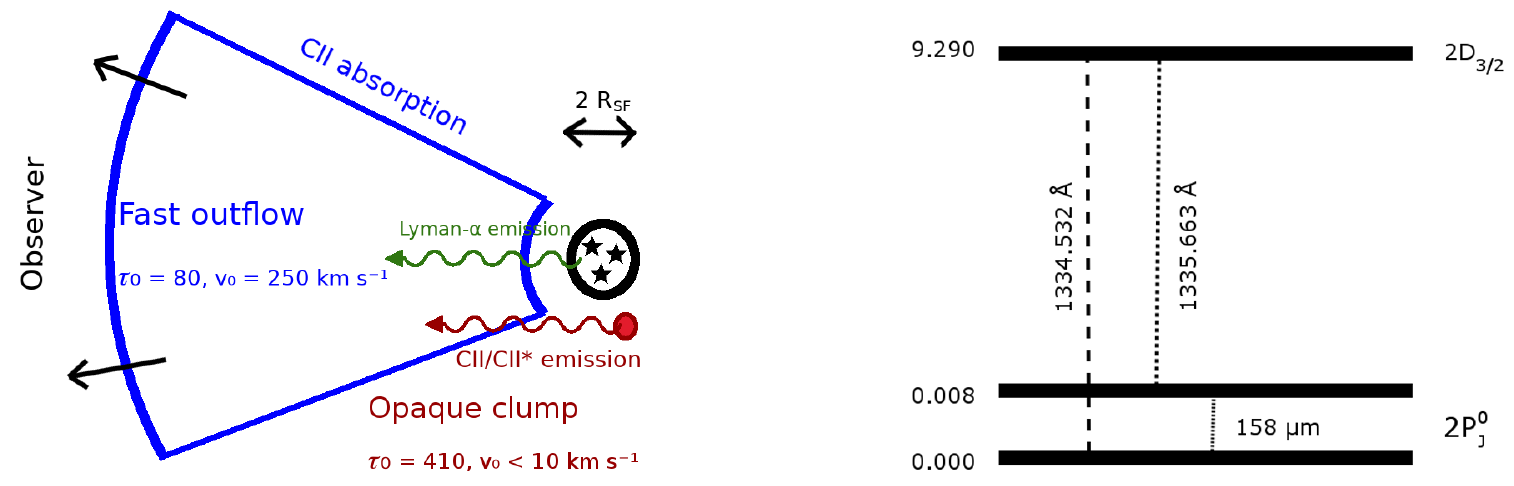}
    \caption{{\textit{Right}}: Selected energy levels of C+. Energies on the left-hand side are given in eV. {\textit{Left}}: Best-fit two-component model of the outflow structure, showing a fast outflow towards the observer causing \ctwo\, absorption (blue) and an opaque clump located closer to the star-forming region (of radius $R_\text{SF}$) giving rise to \ctwo\, and \ctwo* emission (red).}
\end{figure*}


Blueshifted metal absorption is a widespread tracer of galactic outflow, thought to arise from line-of-sight outflowing absorbing material. In addition to this absorption, strong redshifted resonant \textit{emission} is sometimes seen. In the pioneering model of \citet{Rubin11}, this component indicates symmetric and somewhat opaque outflows, with the outflowing material located behind the galaxy re-emitting scattered photons into the line-of-sight. This gives rise to the `P-Cygni'-like profiles seen in galactic absorption lines such as Na~{\small{I}} D 5891/5897\AA \ \citep{Phillips93,Rupke05,Chen10}, \magtwo \ 2880\AA \ \citep{Weiner09, Henry18} and to some extent Lyman-$\alpha$ (Ly$\alpha$, \citealt{Verhamme15}).

In addition to this back-scattered emission, resonantly trapped radiation sometimes has access to alternative fluorescent decay channels. This occurs when excited atoms decay into an excited ground state, giving rise to emission of a different wavelength to the resonant absorption by the gas \citep{Prochaska11}. The most studied examples of this phenomenon are the fluorescent Si~{\small{II}}* 1309, 1533, 1817 \AA \ emission lines produced after absorption of the resonant Si~{\small{II}} 1304, 1526, 1808 \AA \ transitions \citep{Shapley03, France10, Tucker12} and the plethora of optical Fe~{\small{II}}* fluorescent transitions (e.g. \citealt{Martin12}) which are plentiful enough to enable direct mapping and kinematic studies of opaque outflows \citep{Finley17}. 

The detection of fluorescent channels in other optical lines, such as O~{\small{I}} 1302\AA \ and C~{\small{II}} 1334\AA, is more complicated due to the very small wavelength spacing between the resonant and fluorescent channels, 4\AA \ and 1\AA \ respectively. This leads to the blending of the two effects discussed above, i.e. redshifted resonant emission and fluorescent emission, which are not observationally separated even when excess emission is seen (e.g. \citealt{Rivera-Thorsen15, Rigby18}). These alternative decay channels are however extremely interesting for the study of high-redshift galaxies, as the decays 
\comm{from the excited to the non-excited }
ground states give rise to far-infrared [O~{\small{I}}] 63$\mu$m / [O~{\small{I}}] 145$\mu$m and [C~{\small{II}}] 158 $\mu$m emission lines, which are crucial to the study of early galaxies. 

[C~{\small{II}}], in particular, has been the subject of much scrutiny at $z>5$ since it was discovered that the peak of galactic [C~{\small{II}}] emission is usually offset from the peak of star formation as traced by UV emission \citep{Maiolino15,Jones17,Carniani17}. The emission is also frequently clumpy \citep{Carniani18}. This irregular morphology has been interpreted as a telltale sign of mergers, gas accretion, ionising potential inhomogeneities and/or feedback partially clearing the ISM, which could create ionising escape channels (\citealt{Vallini15,Katz17}). Sensitive complementary tracers, such as C~{\small{II}}* scattering and fluorescence, can refine and enhance our understanding of these complex multi-phase early galactic environments.

In this letter, we report the first unambiguous detection of C~{\small{II}}* 1335 \AA \ emission, a fluorescent decay of the resonant C~{\small{II}} 1334 \AA \ transition (Figure 1), which is also detected in this object. In Section 2 we present our observations and reduction procedures. We then analyse this emission by comparing to a simple semi-analytic line transfer model (SALT) inspired by \citet{Scarlata15} \comm{(thereafter SP15)} in Section 3, and summarise our findings and future prospects for the  C~{\small{II}}* 1335 \AA \ line in Section 4. All magnitudes are quoted in AB system.

\begin{table}
    \centering
    \begin{tabular}{l c c c c r}
    \hline
    \hline
Ion & $\lambda$ & $A_{ki}$ & $f_{ki}$ & $E_i-E_k$ & $J_i-J_k$ \\
 & (\AA) & ($\text{s}^{-1})$ &  & (eV) &  \\
 \hline
 C{\small{ II}} & 1334.532 & 2.41E8 & 0.129 & 0.00 - 9.29 & 1/2 - 3/2 \\
  & 1335.663 & 4.76E7 & 0.0127 & 0.00786 - 9.29 & 3/2 - 3/2 \\
  \hline
    \end{tabular}
    \caption{Relevant atomic data for the ionised states considered here \citep{Tachiev00}.}
    \label{tab:my_label}
\end{table}

\section{Observations}

Observations were carried out with X-Shooter/VLT \citep{XSHOOTER} in Visitor mode between August 31 and September 3$^{rd}$ 2017 (ID : 099.A-0128, PI: R. Ellis). The main goal was to obtain deep high-resolution spectra on two bright $z\sim$6 galaxies selected in the Subaru
High-$z$ Exploration of Low-Luminosity Quasars (SHELLQs) by \citet{Matsuoka16}: J0210-0523 ($m_Y$=23.4) and J0125-0555 ($m_Y$=23.6). The pointing of each target was done using a blind offset, although the sources are seen in a short (60s) exposure.  We used the following slits configurations: 1.0''$\times$11'', 0.9''$\times$11'' and 0.9JH''$\times$11'' respectively for the UVB, VIS and NIR arms of the instrument. The nodding mode was applied with the following exposure times: 756s, 819s and 900s respectively in UVB, VIS and NIR arms. One night was lost due to bad weather and we secured 7.5 hrs on each source in good seeing conditions ($<$0.8''). 

The latest version of the X-Shooter ESO Reflex pipeline (v3.2.0 - \citealt{ESOReflex})  was used to reduce individual exposure. We then combined all data using two different methods: (i) updating the Reflex script to combine all exposures into a final deep exposure and (ii) combining all the individual reduced exposure with several IRAF task.  The two methods give similar results in terms of depth of the spectra, but the first method allows a better estimated of the noise. In the following, the analysis will be done on the spectra combined using \textit{Reflex} scripts. Spectra were extracted using \textit{apextract} routine of \textit{IRAF} within an aperture of 2$\times$seeing.

\begin{figure*}
	\includegraphics[width=1.0\textwidth]{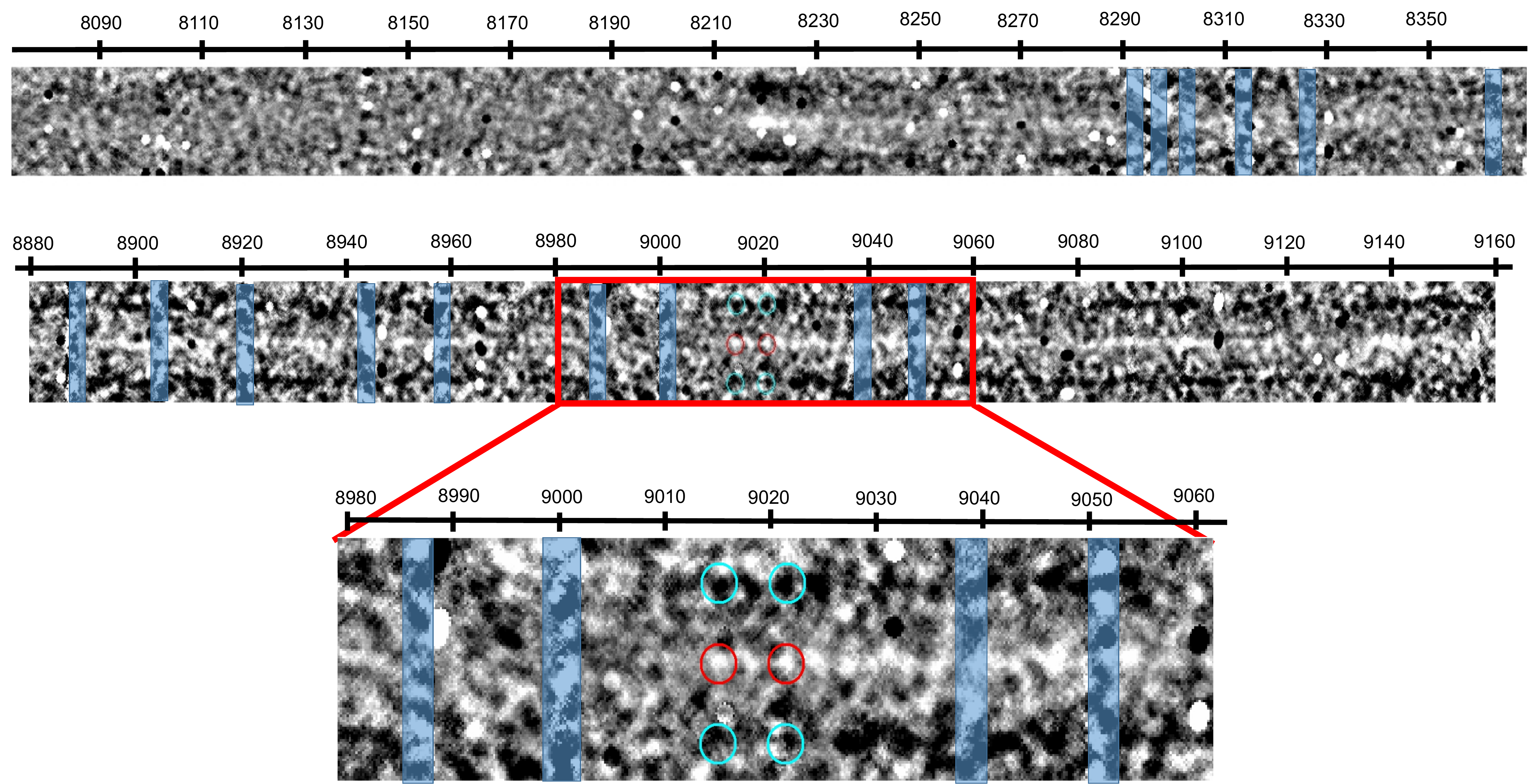}
    \caption{X-Shooter 2D spectrum of J0215-0555. The upper panel shows a cut around Ly-$\alpha$ where the redward continuum is clearly visible. The middle panel shows the spectrum at the position of \ctwo*. The \ctwo\ absorption is clearly seen around $\lambda\sim9000$\AA. The bottom panel is a zoom of the two \ctwo* fluorescent lines (circled in red with the counter-images in blue). Areas affected by sky lines are shaded in blue. }
    \label{fig:raw}
\end{figure*}

\begin{figure*}
	\includegraphics[width=0.7 \textwidth]{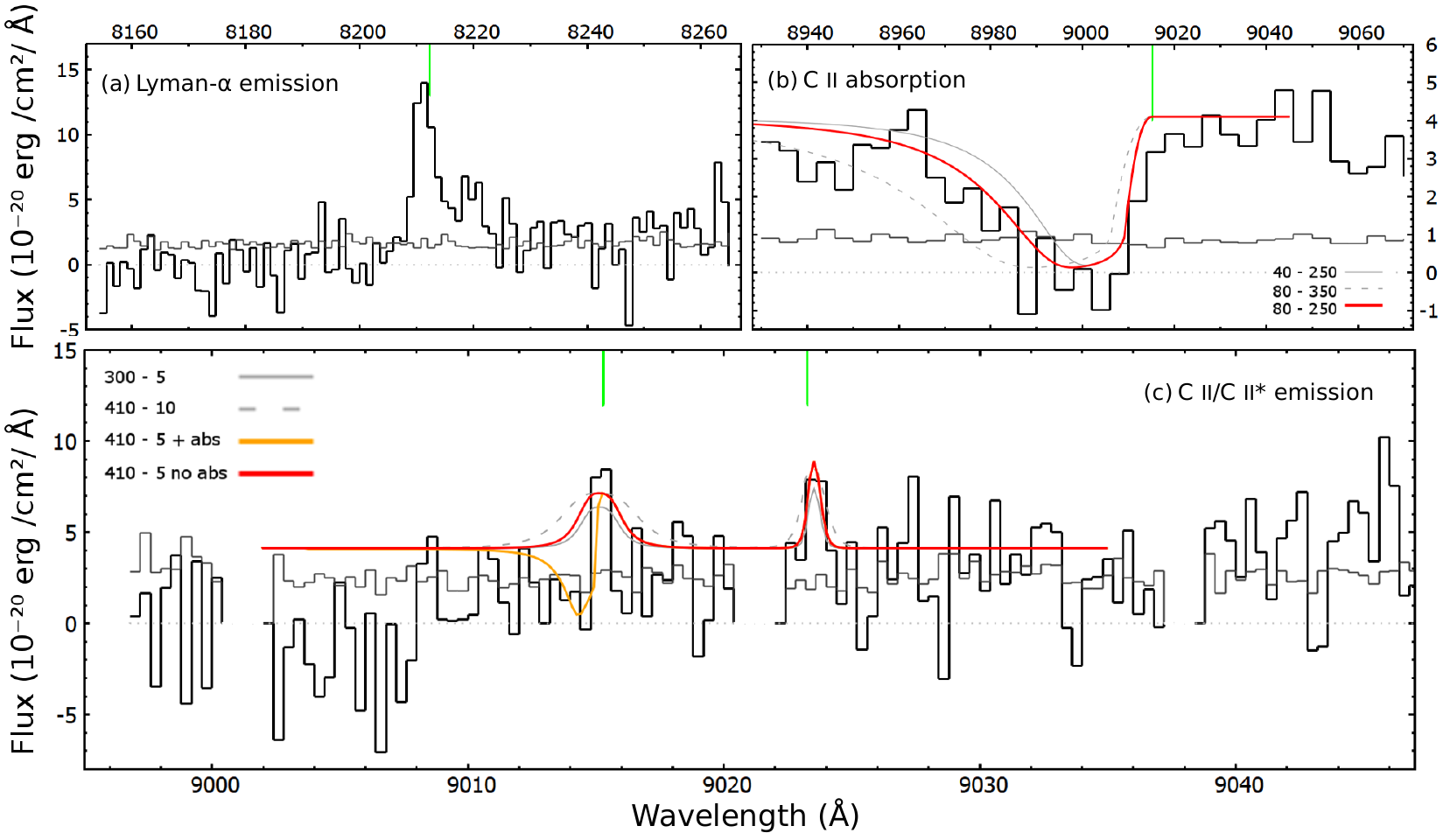}
    \caption{Extracted 1D spectrum of the object. Models are shown as red, orange and grey curves, labelled by the two parameters $\tau_0$ and $v_0$ (see text). The green lines show the positions of the Ly$\alpha$, \ctwo\, and \ctwo* emission lines at a redshift $z=5.7552$.}
    \label{fig:extracted}
\end{figure*}

We detect the \comm{Ly$\alpha$} emission line of galaxy J0215-0555 at $\sim10 \sigma$ above noise (Figure 2, panel a). The full width at half maximum of the line is $\sim 10$ \AA \ (observed) corresponding to $\sim 360$ km s$^{-1}$ velocity width and $z_{\text{Ly}\alpha} = 5.754 \pm 0.002$. Optical continuum emission from the object is detected over the wavelength range $8220<\lambda<9300$\AA, with a median flux of $(4.06\pm0.03)\cdot10^{-20}$ erg s$^{-1}$ cm$^{-2}$ \AA$^{-1}$.
There appear to be intrinsic fluctuations in the continuum emission of $\sim15\%$ across this range.
The continuum emission is interrupted by broad associated metal absorption features including \ctwo 1334 \AA \ (Figure 2, panel b). 
Immediately redward of this absorption, two emission lines are detected at $2.4\sigma$ and $2.7\sigma$ significance, respectively (Figure 2, panels b and c). The spacing of these lines, and alignment with $z_{\text{Ly}\alpha}$, identifies them as \ctwo\, 1334.532 and \ctwo* 1335.663 \AA. If we use a filter consisting of two identical Gaussians representing the spacing of the \ctwo/\ctwo* lines with the FWHM of marginally resolved features (0.6\AA), the significance of the doublet feature rises to $3.3\sigma$. \comm{To further evaluate the significance of the detections, we search the spectrum for all features as bright as the \ctwo* line, which extend over 10 adjacent pixels with consistent (dithered) counter-images. We only find 3 more such candidate emission lines, including the Ly$\alpha$ line.}

The observed velocity spacing of the lines is $\delta v = 268 \pm 16$ km s$^{-1}$, with the uncertainty being due to seeing. This is consistent with the theoretical prediction of  $\delta v = 254.2$ km s$^{-1}$ within $1\sigma$.
\comm{Searching the entire spectrum for pairs of similarly-bright features consistent with this spacing yields no other matches. This considerably strengthens} the identification of the lines \comm{as the probability of this pair coinciding with the Ly$\alpha$ emission line even allowing for an offset of 60\AA\  is less than $5\times10^{-6}$}. 
The derived redshift from double-Gaussian fitting is $z_\text{C II} = 5.7552 \pm 0.0006$, in agreement with Ly$\alpha$.
The FWHM of the lines, of FWHM$=0.5 \pm 0.2$\AA, is consistent with their being spectroscopically unresolved or marginally resolved with respect to the seeing during the observations, $\sim0.6$\AA. 
The \cthree] 1906, 1908 \comm{\AA\ }doublet is not detected, giving a weak constraint on the peak flux ratio of \cthree]/\ctwo $<1.3$ at 2$\sigma$.


\section{Modeling}

The combination of \ctwo\, absorption and \ctwo\, and \ctwo* emission opens up new possibilities to study the geometry and ionisation structure of high-redshift objects. In the case of opaque and spherically symmetric winds, the equivalent widths of emission and absorption in scattered transitions are roughly equal in the majority of cases \citep{Prochaska11}. The obvious departure from this equality in this object indicates a more complex structure, potentially related to clumpiness or inhomogeneous ionisation linked to ionisation escape channels. Therefore, we investigate the implications of the observed emission properties on the source's structure using a semi-analytic approach.
The detections' low signal-to-noise ratio does not warrant a particularly complex modelling. Instead, we model the emission following the simplified 3-parameter SALT model of \comm{SP15}, replacing the SiII/SiII* transitions used by those authors by the \ctwo/\ctwo* transitions given in Figure 1 and Table 1.

In this model, a spherical star-forming region of radius $R_{SF}$ is driving an outflowing spherical wind whose speed scales linearly with distance until a maximum radius $R_W$. This simplistic outflow is described by three parameters $\{\tau_0,v_0,v_{\infty} \}$ corresponding to the opacity of the gas at $R_{SF}$, the outflow velocity at $R_{SF}$, and the velocity at $R_W$. The metal absorption then arises from the outflow `in front' of the galaxy, blueshifted with respect to the observer\comm{, which also re-emits into the line of sight}:
\begin{equation}
I_\text{abs} (x) = \int_{\text{max}(x,1)}^{\sqrt{x^2-1}}\text{d}y \frac{1-e^{-\tau_0/y}}{y - \sqrt{y^2-1}} \comm{- \int_{\text{max}(x,1)}^{v_\infty/v_0}\text{d}y \frac{1-e^{-\tau_0/y}}{2y}},
\end{equation}
where $x=v_\text{obs}/v_0$ and $y=v/v_0$ and an additional boundary condition is applied at the outer radius $R_W$.
Meanwhile, the redshifted outflow `behind' the star-forming region scatters absorbed photons back into the line of sight via both the resonant and fluorescent decay channels, scaling as
\begin{equation}
  I_\text{em, Red/Fluo} (x) = \int_{\text{max}(x,1)}^{\sqrt{x^2-1}}\text{d}y F_\text{Res/Fluo}(\tau) \frac{1-e^{-\tau_0/y}}{2y},  
\end{equation}
where $F_\text{Res/Fluo}(\tau)$ is the fraction of photons decaying via the resonant and fluorescent channels respectively.
$F_\text{Fluo}$ grows with increased gas opacity. For full derivations of these relations and details of the model, we refer the reader to  \comm{SP15}. \comm{We neglect the effect of finite spectroscopic aperture, as we find that the wind would need to extend out to $R_W \sim 35$ kpc for the impact to be comparable to the other sources of uncertainty.}

We ran a $\chi^2$ minimisation fit on the set of parameters $\{\tau_0,v_0,v_{\infty} \}$ in the ranges $0-1000$, $0-1000$ km s$^{-1}$, $0-3000$ km s$^{-1}$ respectively. 
To scale the model predictions to our 1D spectrum, we use the average continuum flux in the object as $F_\text{cont} = 4.06 \cdot 10^{-20}$ erg s$^{-1}$ cm$^{-2}$ \AA$^{-1}$ computed over the whole range where continuum emission is detected. Slight intrinsic variations across wavelength of $ \Delta F\lesssim15\%$ account for the slight apparent mismatch of the continuum in Figure 3.
Nevertheless, it quickly became apparent that a single set of parameters was unable to reproduce both the absorption component (Figure 3, panel b) and the emission components (Figure 3, panel c). Indeed, the central velocity $v_0$ of the outflow strongly determines both the width of the emission and absorption features, and their offset from the systemic redshift. The fluorescent emission features are not offset from the Ly$\alpha$ emission of the object (Figure 3, panel a), and are unresolved at the spectrograph's resolution, implying a very low outflow speed $v_0 \leq 18$ km s${^{-1}}$. On the other hand, the absorption feature extends to $\sim 2000$ km s$^{-1}$ from the redshift of the Ly$\alpha$ line and has a full width at half maximum (FWHM) of $\sim 1100$ km s$^{-1}$.

This velocity mismatch suggests asymmetric outflow.
\comm{We therefore} fit the absorption and emission components separately, under the assumption that they may originate from different parts of the outflow (Figure 1, left). 
\comm{If the outflow is asymmetric, only the line-of-sight velocity difference between  components can be determined in the absence of a known systemic redshift. Because of IGM absorption of Ly$\alpha$, this is unknown and all systemic redshifts consistent with the \ctwo \ absorption trough are permitted by the data ($\Delta z \sim 0.02$). However, these would imply even more complex morphologies including a non-outflowing absorption component. For simplicity, we limit ourselves to outflow-only models and assume $z_{\text{Ly}\alpha}$ as the redshift of star formation.}

The absorption component is best fit by a fast, \comm{opaque} outflow with $v_0 = 250 \pm 50$ km s$^{-1}$, $\tau_0 = 80 \pm 20$, $v_\infty \geq 1700$ km s$^{-1}$ with $1\sigma$ level confidence intervals. The lower limit on $v_\infty$ is due to the fact that the inner regions of the outflow contribute more to profile shape, while we lack the resolution to distinguish the effect further than 2000 km s$^{-1}$ from the object.
Both the resonant and the fluorescent emission components are instead best fit by \comm{an even more} opaque region consistent with no outflow, with $v_0 \leq 10$ km s$^{-1}$, $\tau_0 = 410 \pm 50$, $v_\infty$ unconstrained. The constraint on $\tau_0$ is driven by the line ratio \ctwo/\ctwo*, as higher opacity increases the total emission in both lines but reduces their ratio. We find that the line ratio is consistent with the line strengths at the $1\sigma$ level.

\section{Discussion}

These results suggest that the \ctwo\, absorption originates from an asymmetric fast outflow, as the lack of associated broad scattered light implies it is located preferentially towards the observer and does not fully extend to the far side of the object (Figure 1, left). This conclusion is however somewhat model-dependent, as more detailed studies have shown that not all symmetric outflows display scattered light \comm{for example in the presence of extremely extended ($>35$kpc) outflows which could have been missed} \citep{Prochaska11}. Meanwhile, \comm{t}he \ctwo/\ctwo* scattered light appears to originate in a \comm{more} opaque clump located $\lesssim 50$ km s$^{-1}$ from the location of peak Ly$\alpha$ emission. \comm{If $z_{\text{Ly}\alpha}$ traces star formation, the t}he lack of associated absorption in either \ctwo\, or Ly$\alpha$ implies this clump is located off-centre from the bulk of starlight with respect to the observer but could still be located in the outskirts of the star-forming region.
If these emission lines are thermally broadened, the relation $b=\sqrt{2kT/m}$ implies a temperature $T\lesssim68000$ K, still consistent with photo-ionisation. Finally, using the scaling relation for linear outflows from \comm{SP15}
\begin{equation}
    \tau_0 = \frac{\pi e^2}{m c} f_{ki} \lambda_{ki} n_0 \frac{R_\text{SF}}{v_0},
\end{equation}
and arbitrarily fixing the radius of star formation at $R_\text{SF}=500$ parsec, gives a gas density $n_0 = 1.0 \pm 0.2$ cm$^{-3}$ in the direction of the absorption component, and $n_0 \leq 0.25$ cm$^{-3}$ in the direction of the emission component. 

These rough calculations paint an image of fast, asymmetric winds ($\Delta v \gtrsim 240$ km s$^{-1}$ between the two components) which is entraining gas from the star-forming region ($n_0$ higher by a factor of $\sim4$) and creating a region of lower opacity in the direction of the observer ($\tau$ reduced by a factor of $\sim 5$). In addition, the dense scattering component has a very small velocity width ($v\lesssim10$ km \comm{s$^{-1}$}). Together, these considerations point to a morphology both asymmetric and clumpy -- features which are common in high-redshift galaxies (e.g. \citealt{Carniani18} and therein). The ionisation structure is more challenging to deduce, since no Ly$\alpha$ emission is seen at the location of the fast wind. Instead, the peak of Ly$\alpha$ emission lines up closely with the redshift of the scattering clump, suggesting that all visible Ly$\alpha$ radiation is back-scattered while the `blue peak' is absorbed in the line-of-sight wind. However, since this wind is not symmetric, we speculate than ionisating radiation directed in the opposite direction from the line-of-sight would encounter less absorption. An ionising escape channel could thus potentially exist behind the object.
\section{Summary}We have obtained a 7.5h X-Shooter spectrum of the Lyman-break galaxy J0215-0555 at $z=5.755$ and identified for the first time the decays of \ctwo\, via both the resonant and fluorescent channels, in the same object. While the detection of either line on its own is only significant to 2.4$\sigma$ and 2.7$\sigma$ above the noise level (or 3.3$\sigma$ if taken together), credence is lent to the discovery by \textbf{(i)} the spacing of the lines, of $\delta v = 268 \pm 16$ km s$^{-1}$ compared to the theoretical expectation of $\delta v = 254.2$ km s$^{-1}$; \textbf{(ii)} the close correspondence between the redshift of the doublet and Ly$\alpha$ emission in this object; and \textbf{(iii)} the \ctwo/\ctwo* line ratio, consistent with the observed line strengths within $1 \sigma$. The FWHM of the lines, of FWHM$=0.5 \pm 0.2$\AA, is consistent with them being spectroscopically unresolved with respect to the seeing during the observations, $\sim0.6$\AA. 

In addition to these \ctwo/\ctwo* emission lines, this object displays broad \ctwo\, absorption blueshifted from systemic redshift. We modelled both these components using a simple spherical outflow model, but the narrow width of the emission features only permits extremely low central outflow speeds, of $v_0\leq10$ km s$^{-1}$, while the broad absorption feature extends up to $\sim 2000$ km s$^{-1}$ from the object, and is therefore better fit with higher central outflow speeds, $v_0 = 250\pm50$. Assuming this is indicative of an asymmetric outflow, the strength and ratio of the scattered emission lines is best fit by a highly opaque absorber with $\tau_0=410\pm50$ while the fast outflow has an opacity $\tau_0 = 80 \pm 20$. Using scaling relations, this corresponds to a gas density roughly five times lower in the direction of the dense scattering clump. 

The fluorescent decay \ctwo* 1335\AA \ is difficult to pursue for two reasons: first, it can only be separated from the scattered \ctwo 1334 \AA \ component using high resolution spectroscopy and only when the lines are narrower than their separation; secondly, it only becomes bright at very high opacities ($\tau\sim300$). However, it is of particular interest at high redshift since it directly populates the C+ 2P$^0_{3/2}$ energy level, the source of the forbidden decay [\ctwo] 158 $\mu$m. While a large fraction of this important decay is theorised to come from collisional excitation, fluorescence is an additional channel in the presence of dense outflows. 
The far-infrared [\ctwo] emission line is a crucial tool in the study of $z>5$ galaxies, as it is often extremely bright. In the era of {\textit{JWST}} and the ELT, when infrared spectroscopical observations become easier, the study of fluorescent \ctwo* might become an exciting way to understand the genesis \comm{of} [\ctwo] and close the loop on the cycle of C+.

\section*{Acknowledgements}
We thank Koki Kakiichi and Tucker Jones for their insight. Based on observations collected at the European Organisation for Astronomical Research in the Southern Hemisphere under ESO programme 099.A-0128 (A). SB, NL and RSE acknowledge funding from the European Research Council (ERC) under the European Union's Horizon 2020 research and innovation programme  (grant  agreement No.~669253). This work made us of the NIST Atomic Spectra Database, a facility of the National Institute of Standards and Technology.




\bibliographystyle{mnras}
\bibliography{bibliography} 

\end{document}